\documentclass[journal=nalefd,manuscript=article]{achemso}
\pdfoutput=1

\usepackage{chemformula} 
\usepackage[T1]{fontenc} 
\usepackage{amsmath}
\usepackage{amssymb}
\usepackage{amsfonts}
\usepackage{bm}
\usepackage{soul}
\usepackage{xr}

\author{Nicolas Gastellu}
\author{Michael Kilgour}
\author{Lena Simine}
\email{lena.simine@mcgill.ca}
\affiliation[McGill University]
{Department of Chemistry, McGill University, Montreal, Quebec H3A 0B8, Canada}

\title[Landauer MAC fragments]
  {Electronic Conduction Through Monolayer Amorphous Carbon Nano-Junctions}

\abbreviations{MAC, MO, MCO, PPP, DOS}

\begin{document}

\begin{abstract}
In molecular electronic conduction, exotic lattice morphologies often give rise to exotic behaviors. Among 2D systems, graphene is a notable example. Recently, a stable amorphous version of graphene called Monolayer Amorphous Carbon (MAC) was synthesized. MAC poses a new set of questions regarding the effects of disorder on conduction. In this Letter, we perform ensemble-level computational analysis of the coherent electronic transmission through MAC nano-fragments in search of defining characteristics. Our analysis, relying on a semi-empirical Hamiltonian (Pariser-Parr-Pople) and Landauer theory, showed that states near the Fermi energy ($E_F$) in MAC inherit partial characteristics of analogous surface states in graphene nano-fragments. Away from $E_F$, current is carried by a set of delocalized states which transition into a subset of insulating interior states at the band edges. Finally, we also found that quantum interference between the frontier orbitals is a common feature among MAC nano-fragments. 
 
\end{abstract}


\section{Introduction}
Monolayer Amorphous Carbon (MAC) is a novel 2D carbon material that is topologically distinct from graphene \cite{Toh2020, buchner2014topological}. 
Weak electronic conductance measured in mesoscale MAC\cite{Toh2020} is attributed to the disordered nature of this material. Temperature dependent conductance experiments identified charge transport mechanism in mesoscale MAC as rare-chain hopping \cite{Toh2020, rare_chain_hopping}. This mechanism is characteristic of disordered, quasi-one-dimensional materials. At the nanoscale, MAC is a patchwork of fragments that are correlated on length-scales of 2-3nm, as determined by the decay of the atomistic pair-correlation function \cite{pixelcnn}. In part, mesoscale conduction emerges from the properties of the nano-fragments which, in spite of being small in size, sample a rich configuration space. As a precursor to a study of mesoscale conduction, we here set out to find the common and defining electronic features of the nano-scale fragments of MAC. Our analysis will focus on the coherent electronic transmission properties in an ensemble of isolated MAC nano-fragments.

Geometrical diversity within our ensemble leads to variability in electronic spectra and electronic transmission. When investigating a fragment's transmission profile, it is difficult to distinguish between features peculiar to that particular realization and the transmission characteristics inherent to the material itself. Ensemble-level analysis is therefore needed to identify the defining statistical features of the material. Such statistical approaches have been used in the context of topological matter to demonstrate the existence of nontrivial topological phases in amorphous matter\cite{agarwala_shenoy}. Notably, certain properties of disordered systems emerge only at the ensemble level. For example, while individual finite realisations of an amorphous system might not exhibit any particular symmetry, a set of $N$ amorphous structures from the same ensemble will display continuous translational and rotational symmetries in the thermodynamic limit $N\rightarrow\infty$\cite{chaikin_lubensky}.
A recent illustration of the importance of statistical symmetries in amorphous matter is the protection of edge states (regardless of edge orientation) in amorphous topological insulators by continuous rotational symmetry, and the critical scaling of electronic transport in such systems\cite{spring2021}.

\section{Methods}
\begin{figure}
  \includegraphics[width=1.0\textwidth]{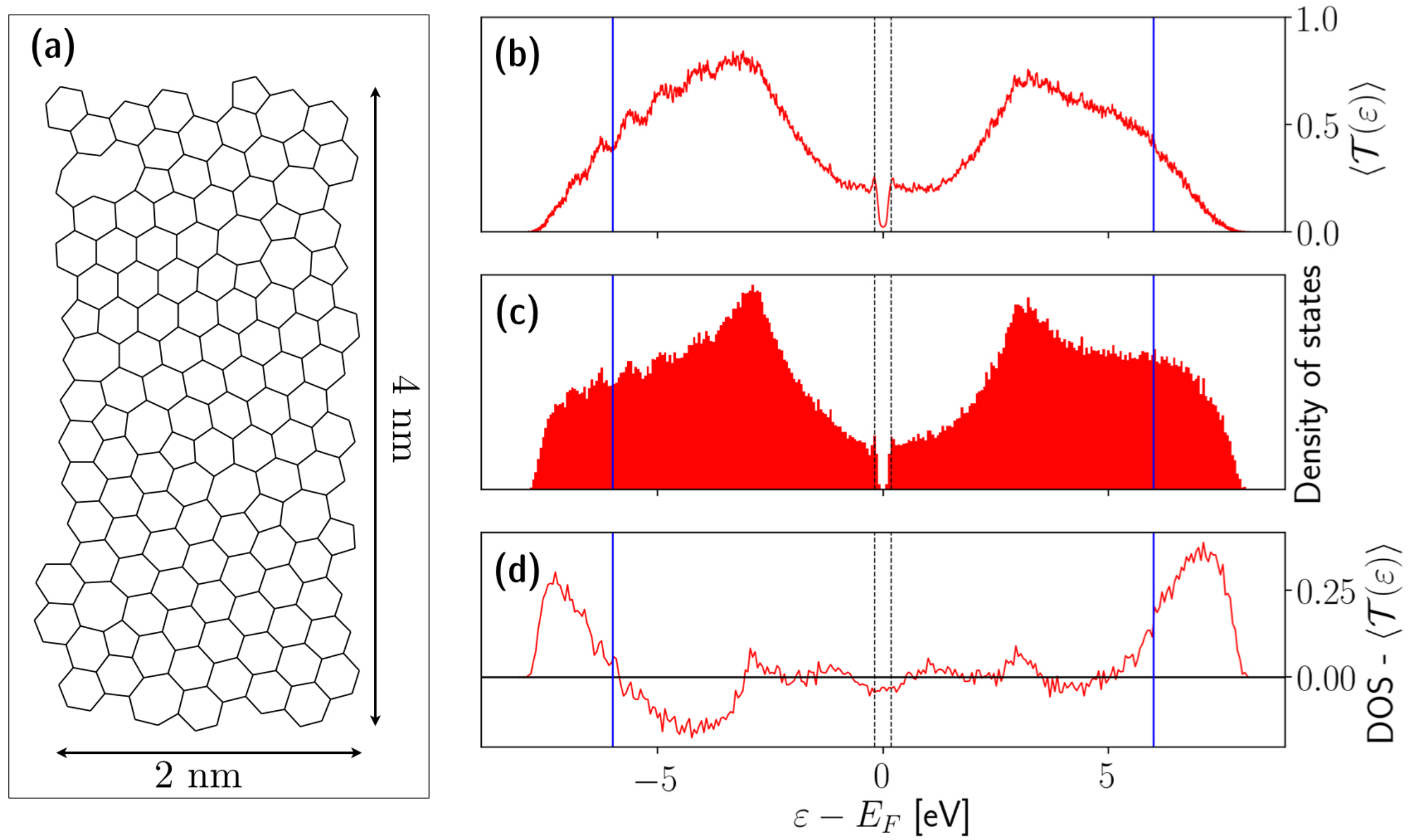}
  \caption{\textbf{(a)} A sample structure of a 2$\,$nm$\,\times\,4\,$nm$\,$ MAC nano-fragment: honeycomb patterns are interrupted by five- and seven-member rings and disordered patches. 
  \textbf{(b)} Ensemble averaged transmission function of MAC nano-fragments (413 samples were used). 
  \textbf{(c)} Density of states (DOS) of the ensemble of 413 MAC nano-fragments. The vertical dashed lines are placed at $\varepsilon = \pm \langle\Delta\varepsilon\rangle/2$, where $\langle\Delta\varepsilon\rangle$ denotes the average bandgap of the ensemble of MAC structures. Verticle solid lines are placed around $\pm 6\,$eV to indicate separation of insulating states from conducting states. 
  \textbf{(d)} Difference plot between the DOS and transmission function computed following normalization by their respective maxima.}
  \label{fig:ensemble_T_DOS}
\end{figure}

In this paper, we perform ensemble-level analysis of conduction states and transmission properties of metal-MAC-metal nano-junctions with MAC nano-fragments 4nm wide by 2nm long (see Figure \ref{fig:ensemble_T_DOS}a for an illustration). 
The junction is created by connecting electrodes to the long edges of our samples.
We chose the dimensions of our fragments small enough to contain varied atomistic patterns but large enough to be considered 2D with identifiable interior and surface sites. 
We generated an ensemble of 413 MAC nano-fragments using Monte Carlo simulations with the AIREBO reactive force field\cite{airebo} as reported in Ref.\citenum{pixelcnn} (see the Supporting Information there for the precise protocol). 
In our analysis, we will highlight the impact of disorder on MAC's electronic properties by drawing occasional comparisons with a graphene nano-fragment\cite{zgnr_edges} of similar size. 
While MAC is topologically distinct from graphene in terms of carbon connectivity\cite{buchner2014topological}, we find that its electronic structure retains some of the features observed in graphene, even if they are distorted by the presence of disorder. We will draw analogies between MAC and graphene to sharpen and illustrate our conclusions.

We employ the semi-empirical Pariser-Parr-Pople (PPP) Hamiltonian\cite{pople1953,pariser_parr1953} to model the electronic structure of MAC nano-fragments.
In this model, only the $\pi$-electrons are included in the calculation of the conduction states. 
This is a tried-and-true approximation for modelling conjugated carbon materials\cite{Warshel1972, Simine2017, ppp_gnr_prb1, ppp_gnr_prb2, pahs_review} which reduces the computational burden and permits calculations on samples with thousands of atoms. 
We benchmarked the method against reported results for graphene nano-fragments\cite{ppp_gnr_prb1} and applied it directly to MAC nano-fragments. 
Further details of our computational methodology including model parameters are given in Section S2 of the Supporting Information. 
   
The electronic transmission function for a two-terminal molecular junction was obtained using\cite{datta_txtbk}:
\begin{equation}
\mathcal{T}(\varepsilon) = \text{Tr}\{\Gamma_L\mathcal{G}(\varepsilon)\Gamma_R\mathcal{G}^{\dagger}(\varepsilon)\}\,,
\label{eq_T}
\end{equation}
where $\Gamma_{\alpha}$ ($\alpha\in\{L,R\}$) are the broadening matrices that quantify the coupling between the nano-fragment and the electrode on edge $\alpha$ (applying the wide band approximation, a constant coupling element of $0.1\,$eV between the leads and a MAC fragment's edge atoms was used, and the edge atoms for each sample were found using the protocol described in Section S3 of the Supporting Information); $\mathcal{G}$ is the retarded Green's function of the open quantum system defined by a MAC structure coupled to semi-infinite leads at its left and right edges: ${\mathcal{G}(\varepsilon) = [\varepsilon - \tilde{\mathcal{H}}]^{-1} = \left[\varepsilon - (\mathcal{H} - \text{i}\frac{\Gamma_L + \Gamma_R}{2})\right]^{-1}}$. For a given MAC realization, expanding the trace in Equation (\ref{eq_T}) in the space of its molecular conducting orbitals (MCOs) $\{|{\psi}_j\rangle\}$ which diagonalize $\tilde{\mathcal{H}}$ and their dual basis $\{|\bar{\psi}_j\rangle\}$ yields an expression of the fragment's transmission function in terms of its quantum interference matrix $Q(\varepsilon)$: ${\mathcal{T}(\varepsilon) = \sum_{j,k} Q_{jk}(\varepsilon)}$\cite{latha,gl}, where:
\begin{equation}
Q_{jk}(\varepsilon) = \frac{\langle\psi_j|\Gamma_L|\psi_k\rangle\,\langle\bar{\psi}_k|\Gamma_R|\bar{\psi}_j\rangle}{(\varepsilon-\lambda_k)(\varepsilon-\lambda_j^*)}\,,
\label{eq:Q_jk}
\end{equation}
and $\lambda_j$ denotes the eigenvalue corresponding to the fragment's $j^{\text{th}}$ MCO. The diagonal elements of $Q(\varepsilon)$ correspond to the single-MCO contributions to the sample's transmission function, while off-diagonal elements can be interpreted as interference terms between MCO pairs.

\section{Results and discussion}
To understand the impact of MAC's configurational disorder on its electronic spectrum and coherent conduction characteristics, we conduct an ensemble-level analysis. 
Figure \ref{fig:ensemble_T_DOS}b shows the transmission function averaged over the entire ensemble.
To better interpret the features of the transmission function, we need to examine the density of states which give rise to conduction. 
Figure \ref{fig:ensemble_T_DOS}c displays the electronic density of states (DOS) of all 413 nano-fragments. 
In the trivial case of entirely independent channels transmitting electrons, we expect the transmission and the DOS to be similar. Coherent transmission is normally resonant at the MCO energies, and when transmission channels are separated by energy gaps larger than their hybridization with electrodes, the correspondence between the density of states and transmission is expected to be very good.
Such correspondence is indeed very close for a graphene nano-fragment over the entirety of its spectrum, with a notable exception around the Fermi energy - a region where constructive quantum interference gives rise to a large spike in transmission (see Supporting Information, Figure S1b). 
Deviations, if observed, indicate more complex underlying processes: quantum interference or qualitative changes to the character of electronic states (e.g. localization).

The agreements and the disagreements between the transmission and the DOS will guide our exploration of this ensemble in search of defining features. 
We visualize the deviations from the trivial conduction behavior in MAC nano-fragments by plotting the difference between transmission and the DOS (appropriately scaled by their largest value) in Figure \ref{fig:ensemble_T_DOS}d.
Several key features stand out. 
We see in Figures \ref{fig:ensemble_T_DOS}b-c that the DOS drops abruptly only at the very extremal energies, but transmission begins to taper down earlier, which suggests that the orbitals in these regions of the ensemble's energy spectrum are poor conduction channels. 
This gives rise to positive deviations in the difference plot in Figure \ref{fig:ensemble_T_DOS}d. 

The negative deviations around $\pm4\,$eV in Figure \ref{fig:ensemble_T_DOS}d witness that the ensemble transmission exhibits broader features than the DOS. 
This is indicative of a straightforward (i.e. weak-interference) transmission regime in which resonance occurs at the MCO energies and is broadened by hybridization with the electrodes. 
This broadening allows a given MCO with energy $\varepsilon$ to also contribute to the transmission of MCOs whose energies are close to $\varepsilon$.
The good agreement between transmission and DOS in a broad range about the Fermi energy (for $|\varepsilon-E_F|\le3\,$eV) is witnessed by minor fluctuations around zero in the difference plot.
Finally, we note as a separate feature that the transmission never goes to zero at the Fermi energy, in spite of the absence of states to carry the current. 
In the following discussion, we refer to the ensemble's manifold of occupied (virtual) orbitals as the occupied (virtual) band.



\begin{figure}
  \includegraphics[width=1.0\textwidth]{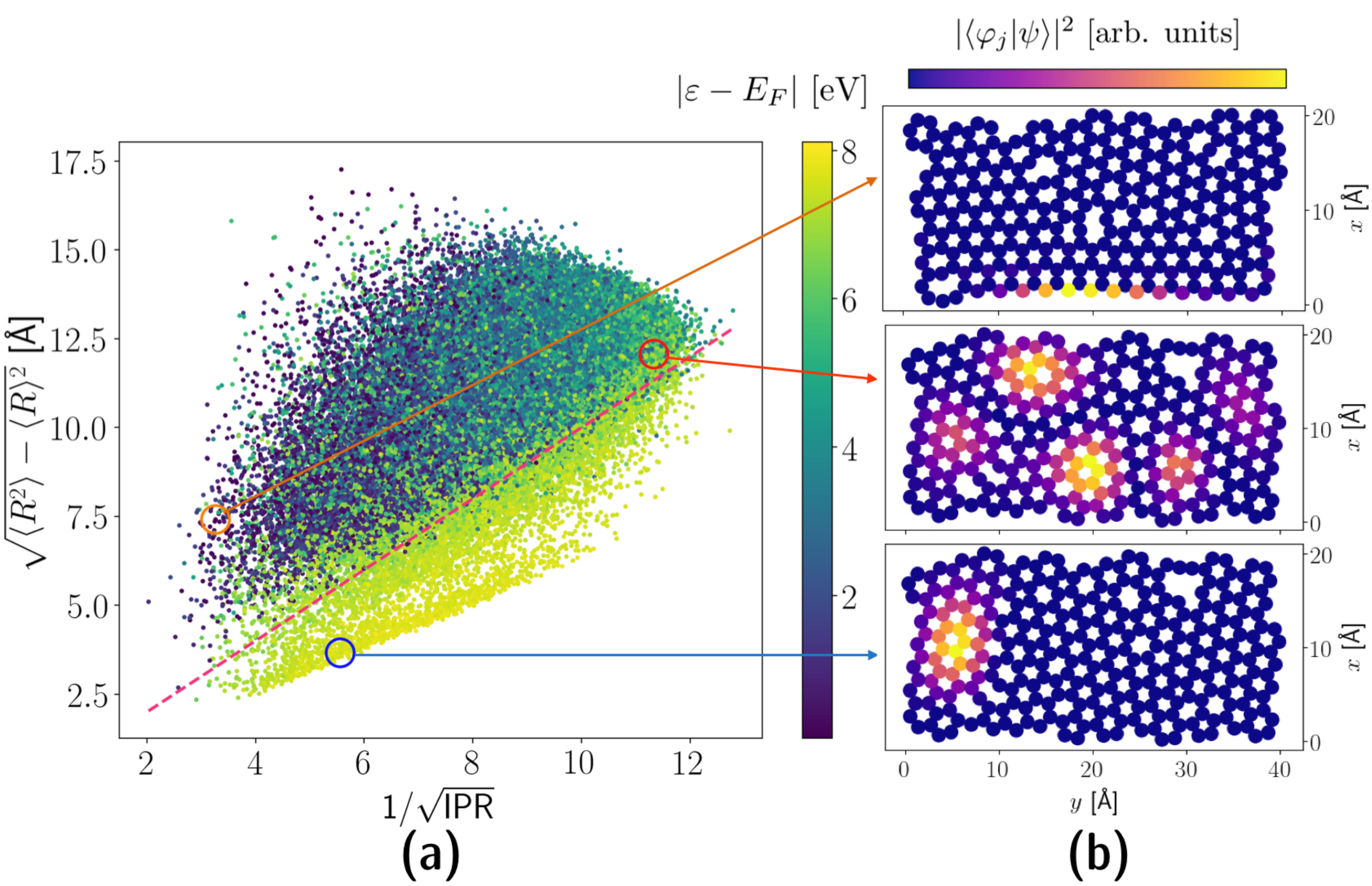}
  \caption{Unraveling the  localization properties of electronic states in MAC nano-fragments. \textbf{(a)} Scatter plot of ensemble MCOs along two different localization metrics: the radius of gyration and the IPR-derived localization length. The dashed line highlights the diagonal along which both metrics are equal. Note the visible separation of near-$E_F$ states (purple) from extremal states (yellow). \textbf{(b)} Sample electronic states represented using their site-space density superimposed on the atomic configuration of a MAC fragment. \textit{Top panel}: A state near the Fermi energy; note its localization on the surface of the MAC structure. \textit{Middle and Bottom panels}: States which lie in the extreme portions of the ensemble's energy spectrum. The MCO shown in the middle panel is a delocalized state (i.e. high ${1/\sqrt{\text{IPR}}}$ and high $R_g$), and the state shown in the bottom panel is compactly localised (i.e. low ${1/\sqrt{\text{IPR}}}$ and low $R_g$).}
  \label{fig:MAC_loc}
\end{figure}

We begin our discussion of these observations from the far edges of the bands. In order to understand what gives rise to the insulating behavior of the ensemble's extremal MCOs, we examine their real-space localization properties. Figure \ref{fig:MAC_loc}a shows the differentiation and clustering of MAC states by scattering them in three dimensions: two different metrics of wavefunction localization, and the orbital energies. We see a clear separation of states at the extremes of the ensemble spectrum (yellow, below the dashed diagonal), from the rest (above the dashed diagonal). 


The origin of the observed energy clustering can be understood in terms of different types of orbtial geometry that our localization metrics capture. Traditionally, one uses the inverse participation ratio (IPR) to distinguish between localised and extended states\cite{bell_dean, wegner}. 
The IPR of electronic state $|\psi\rangle$ is given by: ${\text{IPR}(|\psi\rangle) = \sum_j|\langle\varphi_j|\psi\rangle|^4\,}$, with $|\varphi_j\rangle$ denoting the atomic orbital centered on atom (or site) $j$.
This may be interpreted as a measure of the number of lattice sites where the electronic density of state $|\psi\rangle$ is significant. From the IPR, we can obtain the following localization length of state $|\psi\rangle$: ${1/\sqrt{\text{IPR}(|\psi\rangle)}}$\cite{kramer-mckinnon_review}. 

The IPR localization captures only partial information about the shape of the orbitals and their delocalization in space.
If an electronic state is strongly localized on two opposite edges of a nano-fragment, the IPR metric would miss the delocalization entirely. 
Physical examples of such states include zigzag edge states in graphene nano-fragments, which are localised over few sites on either edge of the nano-fragment, with both edges being separated by possibly very large distances\cite{zgnr_edges, dresselhaus1996edge} (see Figure S2b in the Supporting Information for an illustration). 
In order to capture this delocalization in space, we also compute the radius of gyration for each state given by: ${R_g = \sqrt{\langle R^2 \rangle - \langle R\rangle^2}}$, with: ${\langle R^n\rangle = \sum_j |\bm{R}_j|^n |\langle\varphi_j|\psi\rangle|^2}$ and $\bm{R}_j$ denoting the Cartesian coordinate vector of atom $j$.
This better captures the spatial extent of state $|\psi\rangle$, by evaluating the radius over which it is spread out.

Plotting the states in $(1/\sqrt{\text{IPR}},R_g, E)$ space effectively separates them based on their localization characteristics and in energy. 
The states below the diagonal on Figure \ref{fig:MAC_loc}a tend to be energetically at the band edges (yellow) and also tend to be compactly localized with the radius of gyration lagging behind the IPR localization metric; their spatial extent is roughly commensurate with the number of sites that they occupy (see the bottom panel in Figure \ref{fig:MAC_loc}b for an illustration).
The states above the diagonal tend to be spatially distributed or even `torn', i.e. thinly spread over a small number of distant sites with gaps in between
with most dramatic examples of `torn' states, those having high $R_g$ and low $1/\sqrt{\text{IPR}}$,  resembling the edge states in zigzag graphene nano-fragments localised on two opposite edges of the structure. An an analogous scatter plot of the MCOs of a graphene nano-fragment is shown in Figure S2a in the Supporting Information.

\begin{figure}
  \includegraphics[width=1.0\textwidth]{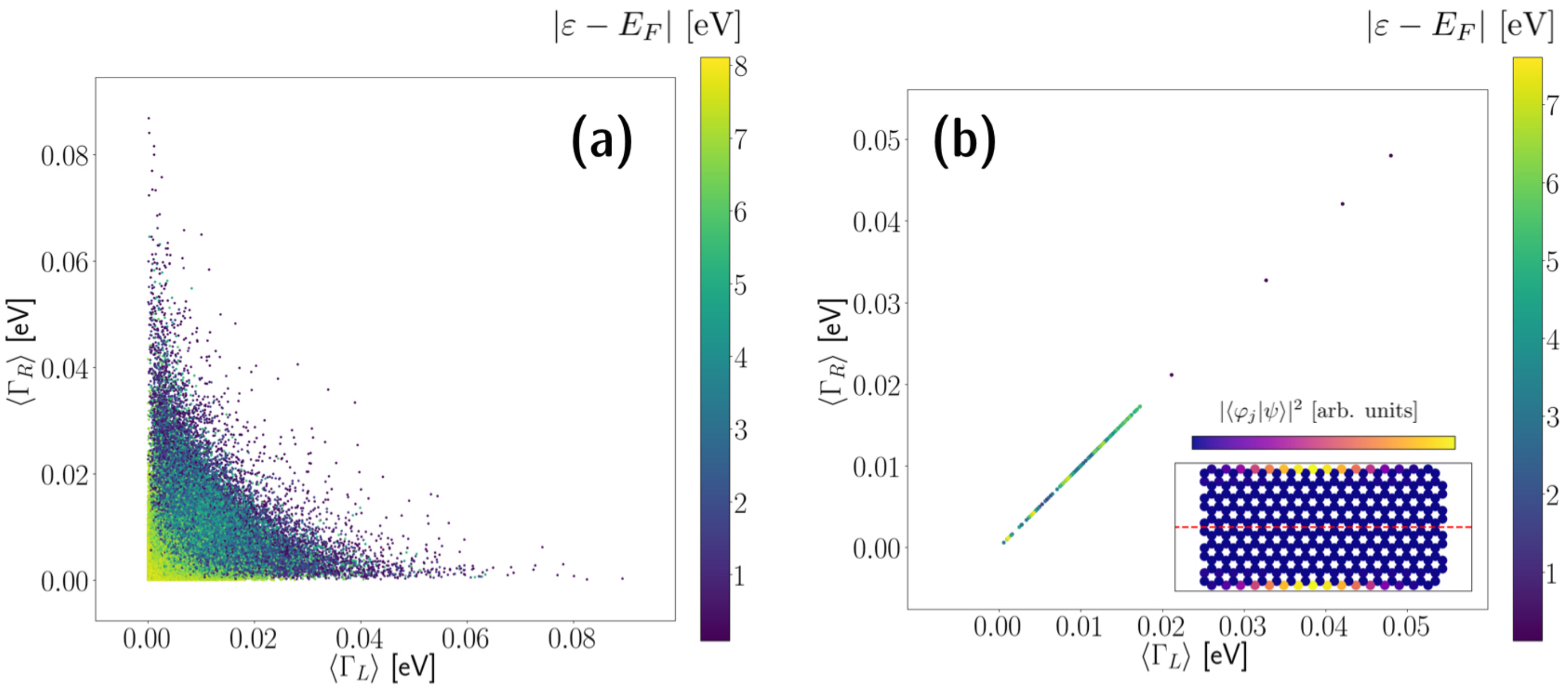}
  \caption{Scatter plot of MCOs' individual coupling to the left and right leads in \textbf{(a)} the MAC nano-fragment ensemble and \textbf{(b)} a graphene nano-flake of similar dimensions. The colour of each point corresponds to the corresponding MCO's distance from Fermi energy $E_F$.
  The inset of \textbf{(b)} shows the HOMO of the graphene nano-flake and the dashed line marks the flake's $\mathbb{Z}_2$ symmetry axis. Note that the HOMO shares the symmetry of the underlying flake, allowing to be simultaneously localized on both edges.}
  \label{fig:gamma_multiplot_final}
\end{figure}

The electronic density maps for a small set of characteristic sample states are visualized in Figure \ref{fig:MAC_loc}b.
The bottom and middle panel depict the wavefunctions of compactly localized states from the extreme portions of the ensemble's energy spectrum.
We note that MCOs with these localisation characteristics tend to be confined to the bulk of the nano-fragment, with little density at the edges.
This leads to weak coupling to the electrode, and consequently low transmission.
Contrastingly, the top panel shows the density of a near-$E_F$ state, which is strongly edge-localised, much like the frontier orbitals in a graphene nano-fragment (compare with the inset of Figure \ref{fig:gamma_multiplot_final}b).
The edge-localised character of the ensemble's near-$E_F$ orbitals leads us to expect them to exhibit better transport properties than the compactly localised states at the extremes of the spectrum.



We further examine the relation between edge localisation and energy by plotting how each MCOs' coupling to the left and right leads vary across the ensemble's energy spectrum in Figure \ref{fig:gamma_multiplot_final}a.
The MCO-lead couplings were computed via a rotation of the coupling matrices $\Gamma_{\alpha}$ into the MCO basis $\{|\psi_n\rangle\}$, and extracting the transformed matrices' diagonal elements $\langle\psi_n|\Gamma_{\alpha}|\psi_n\rangle$.
These expectation values ranged from 0eV, for a completely uncoupled, bulk-localised MCO, to a little under 0.1eV, corresponding to an edge-localised, almost perfectly coupled MCO (like the one depicted in the top panel of Figure \ref{fig:MAC_loc}b).
The clustering of bright yellow points near its origin shows that the extremal sections in the energy spectrum are comprised of states that tend to couple very weakly to the electrodes. This corroborates the conclusion that these are insulating states.
On the other hand, the states with the highest $\langle\Gamma_{L,R}\rangle$ values have energies that are consistently close to $E_F$, which further suggests that MAC hosts robust surface states near $E_F$.


For contrast, in Figure \ref{fig:gamma_multiplot_final}b we show an analogous plot for a graphene nano-fragment 4nm $\times$ 2nm in size (see Figure S1a in the Supporting Information for structure).
For graphene, the correlation between the calculated couplings of conducting states to either electrode is striking but expected \cite{zgnr_edges, dresselhaus1996edge, edges_exptl}. The precise agreement between coupling strengths to either electrode is due to the fact that graphene fragment's edges are related by reflection (or equivalently, rotation by $\pi$) about its $\mathbb{Z}_2$ symmetry axis - the dashed line on the inset of Figure \ref{fig:gamma_multiplot_final}b.
It is also interesting to note that almost every point visible on Figure \ref{fig:gamma_multiplot_final}b is in fact two overlapping data points corresponding to MCOs HOMO$-n$ and LUMO$+n$; their values of $\langle\Gamma_L\rangle$ and $\langle\Gamma_R\rangle$ differ by $\sim 10^{-5}\,$eV, on average in our calculations. MCOs in MAC nano-fragments tend to have a stronger presence on a single edge of the fragment due to broken symmetry.
\begin{figure}
  \includegraphics[width=1.0\textwidth]{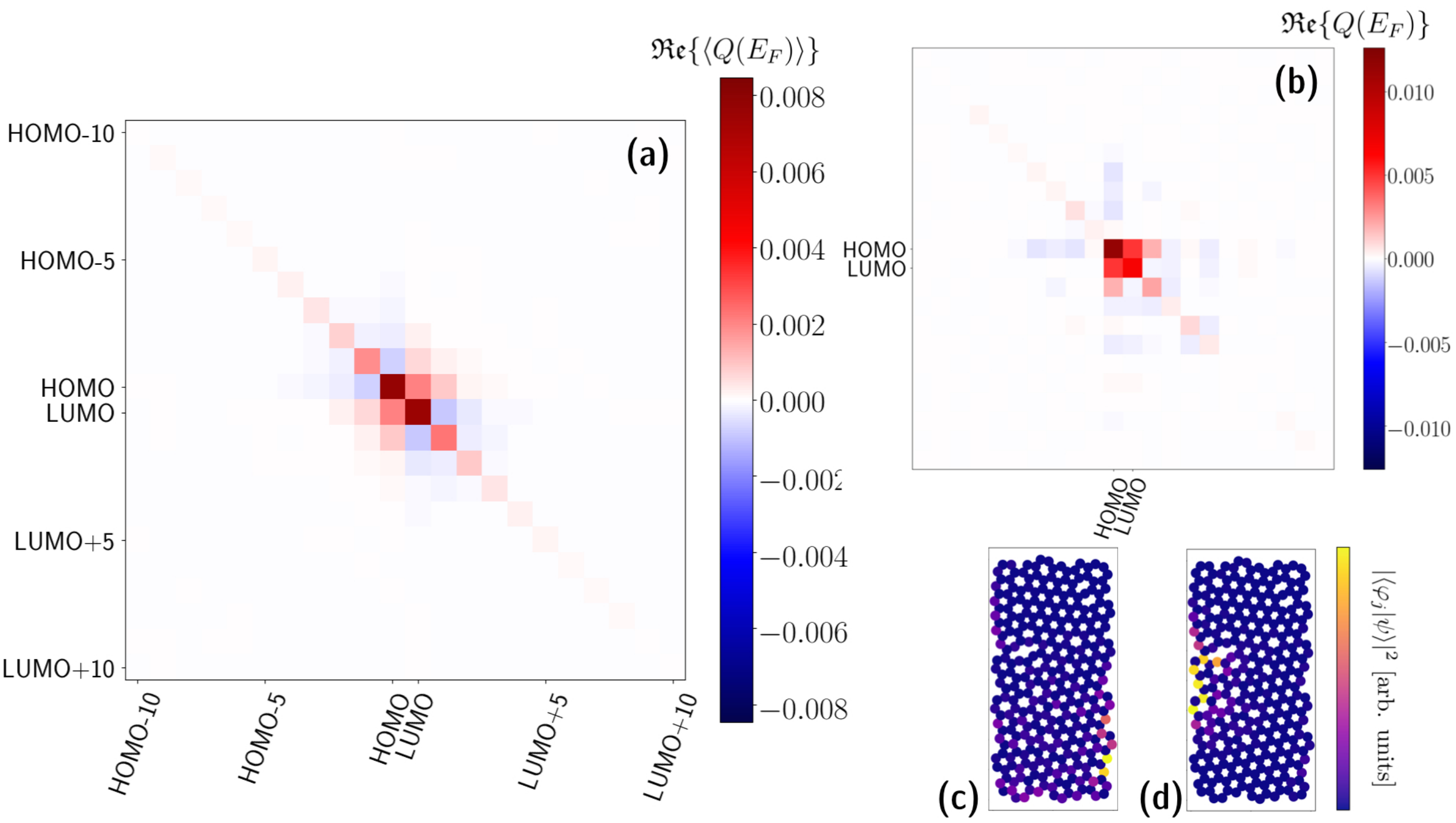}
  \caption{Quantum interference at $E_F$ in MAC. \textbf{(a)} Ensemble-averaged interference matrix. \textbf{(b)} Interference matrix of a single MAC fragment exhibiting strong constructive QI between its frontier orbitals. This MAC fragment's highest occupied molecular conductance orbital and lowest unoccupied molecular conductance orbital are respectively depicted in \textbf{(c)} and \textbf{(d)}. Note that only the matrix elements corresponding to the 10 MCOs closest to $E_F$ in either band are shown in \textbf{(a)} and \textbf{(b)}. All other entries vanish.}
  \label{fig:qi}
\end{figure}

Finally, we discuss the role of constructive quantum interference (QI) in giving rise to the finite transmission within the bandgap (see Figure \ref{fig:ensemble_T_DOS}b-c). 
We demonstrate that interference between frontier orbitals is a defining feature of MAC nano-fragments by observing that interference signal is present in the ensemble-averaged $Q$ matrix defined in Equation (\ref{eq:Q_jk}), evaluated at $E_F$, and plotted in Figure \ref{fig:qi}a. 
We observe weak but positive signal for quantum interference between the frontier and nearby orbitals, which suggests that finite transmission within a given sample's bandgap is due to constructive QI. Notably, the interference between other orbitals averages out to zero at the ensemble level. Figure \ref{fig:qi}b shows the QI matrix of a MAC realization exhibiting particularly pronounced constructive interference between its frontier orbitals at $E_F$.
Figures \ref{fig:qi}c-d depict the frontier conducting orbitals of this MAC structure. 
The interference between these two states is explained by their partial spatial overlap at the edges.


\section{Conclusion}
We have performed ensemble-level analysis of electronic transport properties on monolayer amorphous carbon nano-fragments, and identified several defining features for this disordered nano-material. We found insulating states clustered at the far edges of the occupied and virtual bands and characterized them as compactly localized in the interior of the material. We further found surface states similar to those in graphene near the Fermi energy, and elucidated how lattice disorder, in disrupting the reflection symmetry, confined such states to a single edge suppressing long-ranged electron transport. Finally, we discussed the role of quantum interference in conduction through MAC nano-fragments, finding that constructive interference between frontier orbitals at the Fermi energy emerges as a defining feature of MAC nano-fragments at the ensemble level. 
Future work will be directed at understanding how mesoscale conduction properties in MAC arises from these nano-scale characteristics of its fragments.

\begin{acknowledgement}

Support from NSERC Discovery grant RGPIN-2019-04734 is gratefully acknowledged. Computations were performed on Calcul Quebec supercomputers (Beluga) funded by Canada Foundation for Innovation (CFI). 
\end{acknowledgement}

\begin{suppinfo}

The following files are available free of charge.
\begin{itemize}
  \item Supporting Information: Description of MAC fragment generation method, details on the Pariser-Parr-Pople model and its parametrisation, description of edge-site identification method, and a discussion of the transmission coefficient, MCO geometry, and quantum interference of a 2$\,$nm$\,\times\,$4$\,$nm graphene nano-fragment.%
\end{itemize}

\end{suppinfo}

\bibliography{refs}

\end{document}


\section{S1. Generating the ensemble: simulation of MAC nano-fragments}

We generate the MAC nano-fragments using a Monte Carlo (MC) approach.
Starting from a set of 610 randomly positioned carbon atoms with positions $\bm{R}_n = (x_n, y_n, 0)$, with $x_n$ and $y_n$ independently and sampled from the uniform distribution $\mathcal{U}(0,L)$, and $L=40\,$\AA.
This random initial structure is then relaxed using the AIREBO forcefield\cite{airebo} as implemented in the LAMMPS molecular dynamics (MD) package\cite{lammps}.
Each MC step consists in applying a Stone-Wales (SW) deformation on a pair of neighbouring carbon atoms (defined as being within $r_{\text{max}} = 1.80\,$\AA$\;$ of each other) and relaxing the resulting structure using MD.
This new structure is then accepted with probability $p = \text{min}\left\{1,e^{-\Delta E/k_{\text{B}}T}\right\}$, where $\Delta E$ is the difference between the energies of the new structure and of the structure before the SW deformation, $k_{\text{B}}$ is Boltzmann's constant, and $T$ is the temperature set for the MD simulation.
This operation is then repeated for $N_s$ MC steps.

This procedure is divided into four stages, each characterised by the selection method of which bond to deform, and distinct values of $N_s$ and $T$.
The details of each stage are explained in detail in Ref. \citenum{pixelcnn}.

This MC algorithm yields MAC structures that have 610 atoms and span 40$\,$\AA$\,\times\,$40$\,$\AA. 
We then cleave each generated structure in half along the $y$-axis to obtain our 20$\,$\AA$\,\times\,$40$\,$\AA$\;$MAC nano-fragments.
All monovalent carbon atoms (e.g. at the edges of the nanofragments) are removed before computing a given MAC structure's electronic structure and transmission properties.

\section{S2. Electronic structure: PPP (Pariser-Parr-Pople) Hamiltonian}

We model the electronic structure of each MAC structure using the Pariser-Parr-Pople (PPP) model Hamiltonian\cite{pariser_parr1953,pople1953}.
We assume all carbon atoms in each MAC realization to be $sp^2$ hybridized.
The PPP Hamiltonian explicitly accounts for the $\pi$-electrons only.
Further assuming that each carbon atom contributes a single electron to its MAC fragment's $\pi$ system, our basis set includes one atomic orbital (the $p_z$ orbital) per carbon. 

Solving the PPP model at the Hartree-Fock level of theory, we obtain the following expression for a MAC fragment's electronic energy:
\begin{equation}
    E = \sum_{\mu,\nu} P_{\mu\nu}(H_{\mu\nu} + F_{\mu\nu})\,,
\end{equation}
where $P$, $H$, and $F$ are the MAC molecule's density matrix, core Hamiltonian matrix, and Fock matrix, respectively.

The elements of the density matrix defined as:
\begin{equation}
    P_{\mu\nu} = 2\,\sum_{n}^{\text{occ}} \langle\psi_n|\varphi_{\mu}\rangle\langle\varphi_{\nu}|\psi_{n}\rangle\,,
\end{equation}
with $|\varphi_\mu\rangle$ denoting the $p_z$ orbital centered on carbon atom $\mu$, $|\psi_n\rangle$, the structure's $n^{\text{th}}$ molecular orbital (MO), and the sum running over all occupied MOs.

The core Hamiltonian matrix off-diagonal elements are defined following H\"{u}ckel theory, for $\mu\neq\nu$:
\begin{align}
    H_{\mu\nu} = 
    \begin{cases}
    \beta_{\mu\nu} &\text{for atoms }\mu,\,\nu\text{ nearest neighbours}\\
    0 &\text{  else,} 
    \end{cases}
\end{align}
where sites $\mu$ and $\nu$ are defined as nearest neighbours if they are within 1.8\AA$\;$ of each other (as in our MC generation scheme).
The resonance integrals $\beta_{\mu\nu}$ are evaluated using the Lindberg approximation:
\begin{equation}
    \beta_{\mu\nu} = e^{\zeta(R_{\mu\nu} - R_0)}\,\left(\beta_2 + \beta_2(R_{\mu\nu} - R_0)\right)\,,
\end{equation}
where $\zeta$, $\beta_1$, and $\beta_2$ are semi-empirical parameters describing $sp^2$ carbon, $R_{\mu\nu}$ is the distance between atoms $\mu$ and $\nu$,  and $R_0$ is the equilibrium bond length between two $sp^2$ hybridized carbon atoms (also a semi-empirical parameter of the PPP model).
The planar geometry of the MAC fragments allows us to assume perfect conjugation when evaluating $\beta_{\mu\nu}$.

The diagonal terms of $H$ are given by:
\begin{equation}
    H_{\mu\mu} = \alpha_\mu - \sum_{\rho\neq\mu}\gamma_{\mu\rho}\,,
\end{equation}
where $\alpha_\mu$ represents the valence ionization parameter at the atom $\mu$, and the electron-core repulsion is described by the second term, with $\gamma_{\mu\rho}$ denoting the two-center Coulomb repulsion integral between $|\varphi_{\mu}\rangle$ and $|\varphi_{\rho}\rangle$.

We treat the two-center Coulomb interaction using the Mataga-Nishimoto approximation\cite{mataga_nishimoto1, mataga_nishimoto2}:
\begin{equation}
    \gamma_{\mu\nu} = \frac{|e|^2}{\tilde{\varepsilon}(R_{\mu\nu} + a_{\mu\nu})}\,,
\end{equation}
with $|e|$ denoting the elementary charge, $\tilde{\varepsilon} = 2.7$, the dielectric constant, and:
\begin{equation}
    a_{\mu\nu} = \frac{|e|^2}{\gamma_{\mu\mu} + \gamma_{\nu\nu}}\,.
\end{equation}
The diagonal elements of $\gamma$ also follow from the Mataga-Nishimoto approximation:
\begin{equation}
   \gamma_{\mu\mu} = \frac{I-A}{\tilde{\varepsilon}}\,,
\end{equation}
where $I$ is the valence ionization potential and $A$ is the electron affinity of carbon.

Finally, the elements of the Fock matrix are given by:
\begin{align}
    F_{\mu\nu} &= H_{\mu\nu} - \frac{1}{2}P_{\mu\nu}\gamma_{\mu\nu} \\
    F_{\mu\mu} &= H_{\mu\mu} + \frac{1}{2}P_{\mu\mu}\gamma_{\mu\mu} - \sum_{\rho\neq\mu} P_{\rho\rho}\gamma_{\mu\rho}\,.
\end{align}

The molecular orbitals are obtained using the self-consistent field scheme. 
The parametrization of the PPP Hamiltonian was adopted from previous work.
For a complete list of the numerical values of the parameters used in our calculations, we refer the reader to the supplementary materials of Refs. \citenum{ppp_params1} and \citenum{ppp_params2}.

\section{S3. Identification of edge sites}

The problem of systematically identifying the edge sites of a MAC realization is non-trivial (and somewhat ill-defined), owing to MAC's edge disorder. To this end, we resort to the concave hull algorithm\cite{concave_hull} (a variation of the more well-known convex hull algorithm), which adopts a $k$-nearest neighbours approach to finding the boundary elements of a given two-dimensional point set.
This \textit{ad hoc} procedure is readily automatable (therefore well-suited to ensemble-wide application) and scales linearly with the number of the points in the input set\cite{concave_hull}.

This algorithm returns a list of all carbon atoms on the edges of a given MAC structure.
Taking the  $x$ axis as the transport direction, we then define the left (right) edge as the the set of edge atoms whose $x$ coordinate is within 3\AA$\;$ of the minimum (maximum) $x$ coordinate in the considered structure.
We orient the MAC nano-fragments such that their shorter edges lie along the $x$ axis (see figure 1a in the main text).

Determining the edge atoms in the graphene nanoribbon we simulated was a much simpler task; we defined thenano-fragment's left (right) edge as the set of carbon atoms with the minimum (maximum) $x$ coordinate.

\section{S4. Electronic states and transmission in a graphene nanoribbon}

\begin{figure}
  \includegraphics[width=1.0\textwidth]{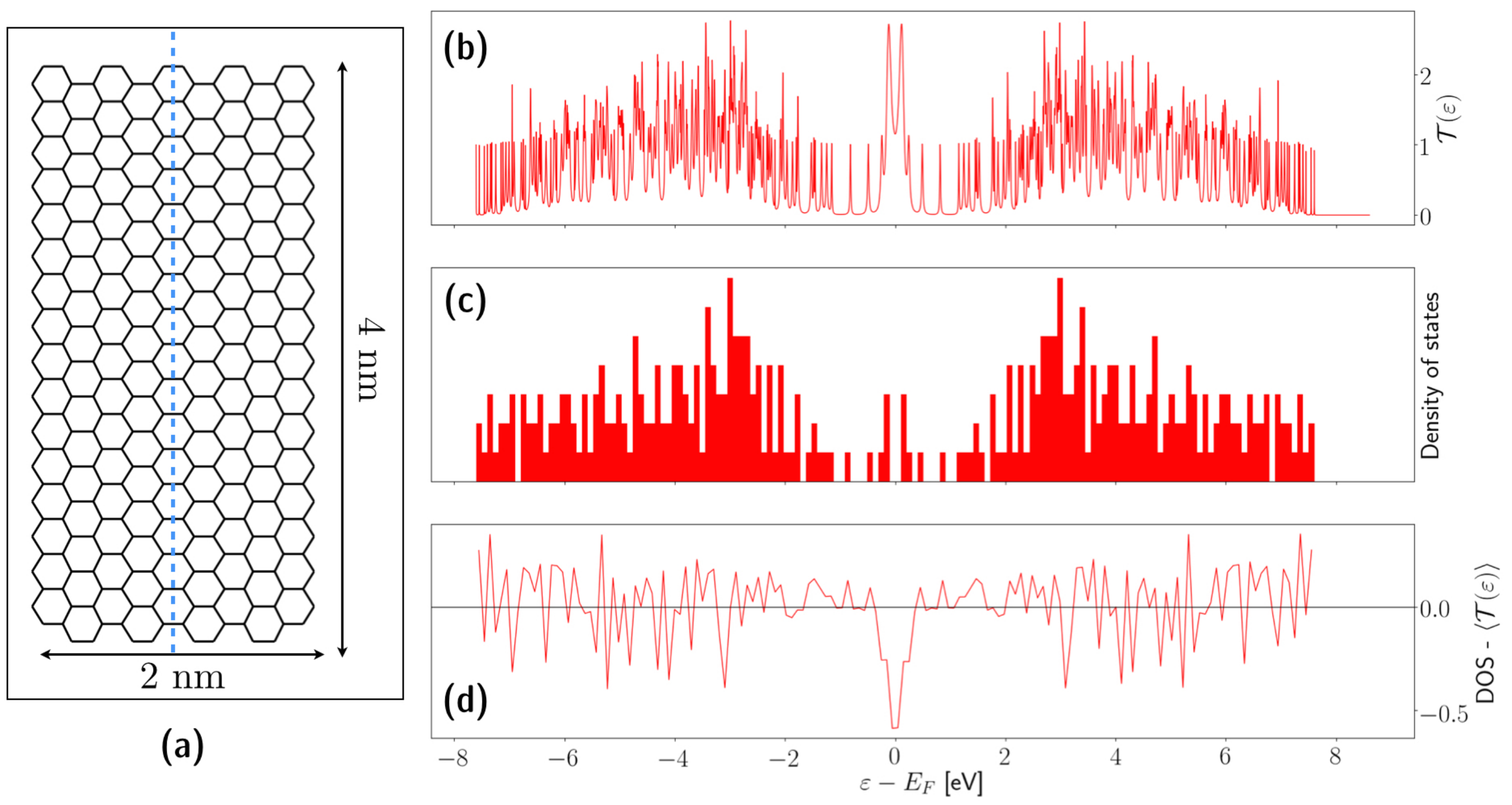}
  \caption{\textbf{(a)} Representation of the 2$\,$nm$\,\times\,4\,$nm$\:$ graphene nanoribbon whose transmission function and DOS is plotted on the other panels. The zigzag edges (along the vertical axis) are coupled to the leads. The blue dotted line marks the nanoribbon's $\mathbb{Z}_2$ symmetry axis.
  \textbf{(b)} Transmission function of the graphene nanoribbon plotted in \textbf{(a)}. Note the strong resonance near $E_F$. 
  \textbf{(c)} Density of states (DOS) of the graphene nanoribbon.
  \textbf{(d)} Difference plot between the DOS and transmission function computed following normalization by their respective maxima.}
  \label{fig:aGNR_T_DOS}
\end{figure}

In this section, we examine the transmission properties and spatial configuration of the molecular orbitals supported by a graphene nanoribbon (shown in Figure \ref{fig:aGNR_T_DOS}a) of similar dimensions as the MAC structures in our ensemble.
The graphene nano-fragment's long edges are of zigzag type, and its short edges are of armchair type.
As in the case of our MAC nano-fragments, we couple the long edges of the graphene structure to the leads.
Comparing the results obtained for this graphene nanoribbon to those obtained for the MAC ensemble (discussed in the main text) helps illustrate how structural disorder gives rise to MAC's transmission properties.

Figure \ref{fig:aGNR_T_DOS}b displays the graphenenano-fragment's transmission coefficient $\mathcal{T}(\varepsilon)$, whose profile matches that of its density of states (shown in Figure \ref{fig:aGNR_T_DOS}c), with a notable exception near its Fermi energy $E_F$.
Ignoring this exception for now, the difference between the graphenenano-fragment's DOS and $\mathcal{T}(\varepsilon)$ (show.n in Figure \ref{fig:aGNR_T_DOS}d) does not display significant structure the  over the rest of thenano-fragment's energy spectrum, which is consistent with the agreement between its DOS and $\mathcal{T}(\varepsilon)$ noted above.
However, this difference curve appears to make more frequent excursions below $\text{DOS}-\mathcal{T}(\varepsilon) = 0$ line than above it, which suggests that thenano-fragment's molecular conductance orbitals' (MCOs) transmission properties are enhanced by constructive quantum interference (QI).
This negative difference is particularly acute near $E_F$, where a relatively small number of MCOs contribute to two significant peaks in the sample's transmission.
We furthermore note that that $\mathcal{T}(\varepsilon)$ remains greater than 1 within thenano-fragment's HOMO-LUMO gap, despite the absence of channels in this range of energies ($|\varepsilon-E_F|\le 0.22\,$eV).
Finally, we also note the symmetry about $E_F$ of graphene's DOS and transmission coefficient, which is a consequence of its particle-hole symmetry (PHS).

\begin{figure}
  \includegraphics[width=1.0\textwidth]{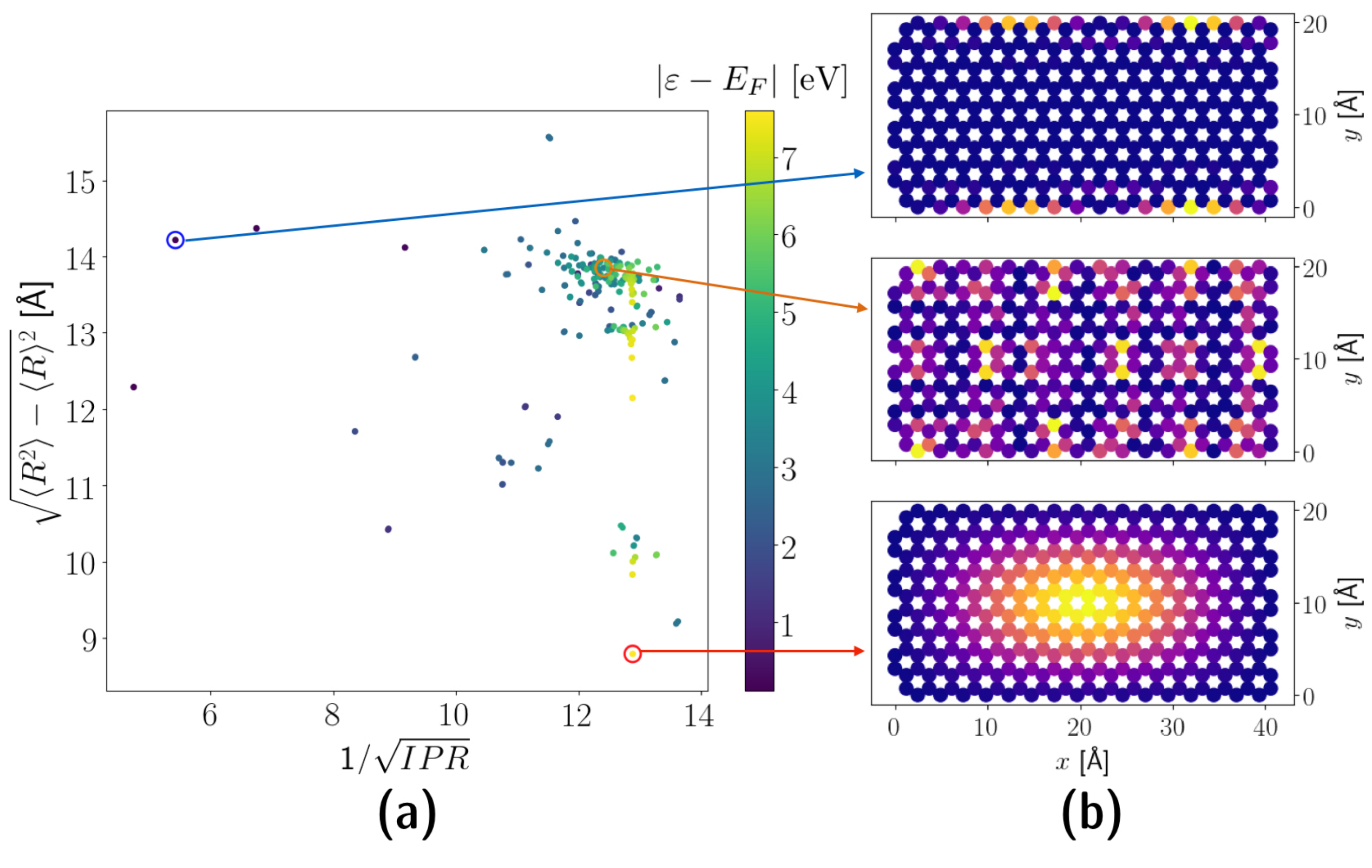}
  \caption{Delocalisation and symmetry of graphene  orbitals. \textbf{(a)} Scatter plot of graphene nanoribbons MCOs along two different localization metrics: the radius of gyration and the IPR-derived localization length. Note the clustering of states in the upper right-hand corner, demonstrating the highly delocalised nature of graphene MCOs. The edge states near $E_F$ stand out in the upper left-hand corner of the scatter plot. \textbf{(b)} \textit{Top panel}: An edge state near the Fermi energy of the nanoribbon. \textit{Middle panel}: Wavefunction of a delolcalised MCO. \textit{Bottom panel}: `Localised' state of the graphene nanoribbon. This MCO covers a significant portion of the structure despite having the lowest radius of gyration. Note the $\mathbb{Z}_2$ symmetry of all MCOs plotted above about thenano-fragment's symmetry axis (roughly the $y\simeq10\,$\AA$\,$ axis, see Figure \ref{fig:aGNR_T_DOS}a).} 
  \label{fig:aGNR_loc}
\end{figure}

The high transmission coefficient of graphene's MCOs can be attributed to their highly delocalised geometry.
As can be seen in Figure \ref{fig:aGNR_loc}a (which is the graphene nanoribbon's analogous plot to Figure 2a in the main text), most of the graphene nanoribbon's MCOs exhibit a high radius of gyration and are distributed over a large number of lattice sites.
The extended nature of these MCOs' wavefunctions arises from graphene's $\mathbb{Z}_2$ symmetry axis (depicted as a dashed blue line on Figure \ref{fig:aGNR_T_DOS}a).

We also observe that MCOs near $E_F$ (dark purple) display significantly different localisation characteristics than the rest of thenano-fragment's electronic states.
Indeed, these MCOs display the same discrepancy between their two localisation metrics as the near-$E_F$ states in the MAC ensemble (cf. Figure 2a in the main text): a high radius of gyration ($R_g$), but a low inverse-participation-ratio-derived metric ($1/\sqrt{\text{IPR}}$).
These localisation charactersitics appear to correspond to zigzag edge states\cite{zgnr_edges} (see the top panel of Figure \ref{fig:aGNR_loc}b), who are distributed over a handful sites located on opposite ends of the nanoribbon.

The edge-localisation of thenano-fragment's near-$E_F$ states is further confirmed by the fact that they exhibit the strongest coupling to the leads, as can be seen in Figure 3b of the main text.
The peaks in transmission near $E_F$ can therefore be assigned to these edge states, whose geometry allows them to couple very strongly to both electrodes at once and thus efficiently transmit charge carriers across the graphene nanoribbon.
Furthermore, the extended nature of thenano-fragment's other MCOs increases the likelihood that they will overlap at its edges, which in turn makes it more probable that they will interfere constructively.

We therefore note that MAC's structural disorder tends to localise its MCOs.
As can be seen on Figure 2a of the main text, the MAC ensemble's MCOs do not cluster in the upper right-hand corner of the plot; they are instead scattered over a much larger region of $(1/\sqrt{\text{IPR}}, R_g)$ space.
Furthermore, the more localised MAC MCOs tend to be bulk-localised which hinders their ability to transport charge.
This disorder-induced localisation is also manifest in the MAC MCOs that are close to $E_F$, which are usually confined to a single edge edge and are therefore less effective transmission channels than the graphene nanoribbon's edge states.

\begin{figure}
  \includegraphics[width=1.0\textwidth]{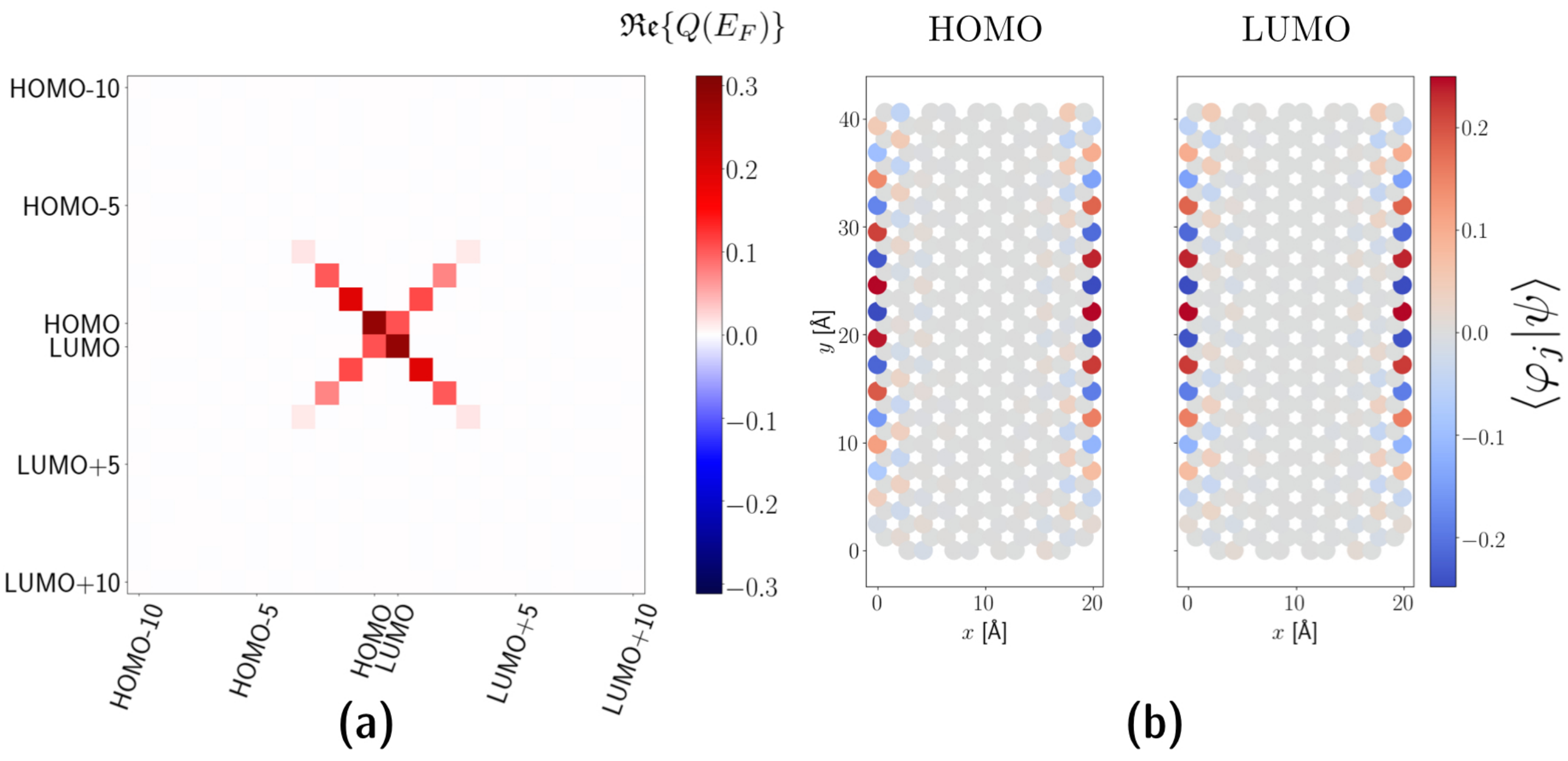}
  \caption{\textbf{(a)} Quantum interference matrix of the graphene nanoribbon at the Fermi energy. Note that only the matrix elements corresponding to the 10 MCOs closest to $E_F$ in either band are shown here. All other entries are zero. \textbf{(b)} Site-space functions of the MAC fragment's HOMO and LUMO. Note that the two wavefunctions only differ by the relative signs of their probability amplitudes on the structure's left and right edges. For the HOMO, both edges are perfectly out of phase (i.e. $\langle\varphi_j|\psi_{\text{HOMO}}\rangle = -\langle\varphi_k|\psi_{\text{HOMO}}\rangle$ where sites $j$ and $k$ are each other's reflections about the $\mathbb{Z}_2$ symmetry axis), whereas both edges are perfectly in phase for the LUMO (i.e. $\langle\varphi_j|\psi_{\text{LUMO}}\rangle = \langle\varphi_k|\psi_{\text{LUMO}}\rangle$, with sites $j$ and $k$ defined as above).} 
  \label{fig:aGNR_QI}
\end{figure}

Finally, we plot the graphene nanoribbon's QI matrix evaluated at its Fermi energy, $Q(E_F)$ on Figure \ref{fig:aGNR_QI}a.
We observe strong constructive between the HOMO$-n$ and LUMO$+n$ for $n\in\{0,1,2,3\}$ (about two orders of magnitude greater than the average QI in MAC, cf. Figure 4a in the main text).
The significant QI between these pairs of MCOs is due to their localisation on the fragment's zigzag edges (which are coupled to the leads) and to graphene's PHS.
Indeed, the wavefunctions MCOs that are equidistant from thenano-fragment's Fermi energy (i.e. HOMO$-n$ and LUMO$+n$) differ only by a few relative phases (see Figure \ref{fig:aGNR_QI}b for an example), as a consequence of graphene's PHS.
Thus, these MCO pairs will strongly overlap and, in the case of states near $E_F$, all of this overlap will occur at the zigzag edges ofnano-fragment, which leads to strong QI between them.
We therefore attribute MAC's significant transmission within its HOMO-LUMO gap to the strong constructive QI we observe in Figure \ref{fig:aGNR_QI}a. 

\bibliography{refs}